\documentclass[preprint,nofootinbib]{revtex4}
\usepackage{epsfig}
\usepackage{amssymb}
\newcommand{\CC}{\Lambda}
\newcommand{\rL}{\rho_{\CC}}
\newcommand{\rLo}{\rho_{\CC 0}}
\newcommand{\rD}{\rho_{\rm D}}

\begin{document}
\begin{titlepage}

\begin{center}
\small{This essay received an \textit{Honorable Mention}
from the Gravity Research Foundation (2018)\\ Awards Essays on
Gravitation}
\end{center}
\vspace{1cm}

\vspace{0.5cm}

\begin{center}
{\Large \bf Brans-Dicke gravity: from Higgs physics\\  to (dynamical) dark energy }


\vskip 0.5cm {\large \bf Joan Sol\`a Peracaula}
\end{center}

\begin{quote}
\begin{center}
Departament de F\'isica Qu\`antica i Astrof\'isica, and Institute of Cosmos Sciences, Universitat de Barcelona\\ Av.
Diagonal 647 E-08028 Barcelona, Catalonia, Spain\\
Email: sola@fqa.ub.edu
\end{center}
\end{quote}
\vspace{0.5cm} \centerline{(Submission date: March 31, 2018)} \vspace{0.5cm}
\centerline{\bf Abstract}
\bigskip
The Higgs mechanism is one of the central  pieces of the Standard Model of electroweak interactions and thanks to it we can generate the masses of the elementary particles. Its fundamental origin is nonetheless unknown.  Furthermore, in order to preserve renormalizability we  have to break the  gauge symmetry  spontaneously, what leads to a huge induced cosmological constant incompatible with observations.  It turns out that in the context of generalized Brans-Dicke theories of gravity the Higgs potential structure can be motivated from solutions of the field equations which carry harmless cosmological  vacuum energy. In addition, the late time cosmic evolution effectively appears like an universe filled with mildly evolving dynamical dark energy mimicking quintessence or phantom dark energy.
\vspace{0.3cm}


\end{titlepage}

\pagestyle{plain} \baselineskip 0.75cm

\section{Higgs potential and harmful vacuum energy}

In the $\CC$CDM, or concordance cosmological model, the measured vacuum energy density at present, $\rLo=\CC/(8\pi G)$, is of order
$10^{-47}$ GeV$^4\sim \left(10^{-3} {\rm eV}\right)^4$ in natural
units. Here $\CC$ is the cosmological constant (CC) and $G$ is Newton's constant.
The discovery of the Higgs boson $\phi$ and the measurement of its mass
($M_\phi\simeq 125$ CeV) implies a value of the electroweak (EW) vacuum energy density of order
$|\langle V_\phi\rangle|\sim  M_\phi^2\, v^2\sim  10^8$ GeV$^4$,  where $v={\cal O}(200)$ GeV is the vacuum expectation value of the Higgs doublet of the standard model. The result is some $55$ orders of magnitude higher than $ \rLo$ .  Such phenomenal  mismatch is at the root of the famous  ``CC problem''\,\cite{Weinberg89,CCP1,CCP2,JSPReview2013} in the context of the standard model of particle physics.

Thus, paradoxically, while the famous Englert-Brout-Higgs mechanism\,\cite{EnglertBrout1964,Higgs1964ab}  for spontaneous symmetry breaking (SSB) of the gauge symmetry is very much welcome in particle physics, it nevertheless carries a profound and unsolved enigma in the context of gravity and cosmology which seems to clash violently with observations.
The problem suggests that the vacuum energy in quantum field theory (QFT) may not be quite the same as the cosmological vacuum energy. The two kind of vacua might be conceptually different and in such case there would be no reason to mix them up.   While a clue to this conundrum  does not seem possible in General Relativity (GR), it might emerge from  Brans-Dicke (BD) type gravity.


\section{Generalized Brans-Dicke action and field equations}

Consider the standard BD-action\,\cite{BD}  with  BD-field, $\psi$, coupled to the curvature R:
\begin{eqnarray}
S_{BD}[\psi]=\int d^{4}x\sqrt{-g}\left[\frac12\,R\psi-\frac{\omega}{2\psi}g^{\mu\nu}\partial_{\nu}\psi\partial_{\mu}\psi-\rho_\CC\right]\,,  \label{eq:BDaction}
\end{eqnarray}
where  $\omega$ is the standard  BD-parameter\,\cite{BD}, and we have included a constant $\rho_\CC$.  Its relation with $\rLo$, defined above, will be elucidated later on. Recall that in the limit $\omega\to\infty$ this action is supposed to reproduce Einstein's GR.
In the BD model, the effective gravitational coupling at any time of the cosmic expansion is a quantity slowly varying with the cosmic time $t$, and reads
\begin{equation}\label{eq:Geff}
G_{\rm eff}(t)=\frac{1}{8\pi\psi(t)}.
\end{equation}
Equivalently, the effective value of the (reduced) Planck mass squared at any time is just given by the BD-field:
$\psi(t)=M_P^2(t)$.
At $t=t_0$ (our time) we have $M_P(t_0)\equiv M_P=1/\sqrt{8\pi G}\simeq2.43\times 10^{18}$ GeV, hence $G_{\rm eff}(t_0)=G$
 is the current value of Newton's coupling.


Let us extend non-trivially the above action by invoking a new scalar field, $\phi$, which is non-minimally coupled to gravity (both derivatively and non-derivatively) and also coupled with the BD-field.  We denote its (as yet) unknown potential by  $V(\phi)$.  The suggested new piece of the action is
\begin{eqnarray}
S[\phi,\psi]=\int d^{4}x\sqrt{-g}\left[\xi R\phi^{2}-\frac{1}{2} g^{\mu\nu}\partial_{\mu}\phi\,\partial_{\nu}\phi  +\frac{\varsigma}{\phi^{2}}G_{\mu\nu}\partial^{\mu}\phi\partial^{\nu}\phi +\eta\phi^{2}\psi  -V(\phi)\right]\,.  \label{eq:SMBDaction}
\end{eqnarray}
Here  $G_{\mu\nu}= R_{\mu\nu}-{(1/2)}g_{\mu\nu}R$ is the Einstein tensor and $\xi,\varsigma,\eta$  are dimensionless coefficients.  Derivative interactions ($\varsigma\neq 0$) with gravity  have been considered before in the literature \,\cite{Amendola93,Capozziello99,Elahe2017}. As we shall see, they are crucial for our purposes. While $\psi(t)$ varies very slowly with the cosmic time, $\phi(t)$ can evolve much faster.  To comply with homogeneity and isotropy, none of them should vary with space. The field $\phi$ will effectively behave as the Higgs scalar. The total action that we propose is the sum  $S_{tot}=S_{BD}[\psi]+S[\phi,\psi]+S_m$, where $S_m$ is  the matter action.  For  $\xi=\varsigma=\eta=0$  the scalar field part of the action boils down to the sum of two decoupled actions $S_{BD}[\psi]+S[\phi]$, with $S[\phi]$ minimally coupled to gravity. This case is of course uninteresting.


Let us first explore the early epoch when the electroweak (EW) phase transition occurs.  This epoch is characterized by the scalar field dominance and we can neglect both $\rL$ and  $S_m$.  Within the (flat) FLRW metric ($ds^2=-dt^2+a^2(t)d{\bf x}^2$), the variation of the total action, $S_{tot}$, with respect to the BD-field  $\psi$ yields
\begin{eqnarray}
3\dot{H}+6{H}^{2} - \omega \frac{\ddot{\psi}}{\psi}+\frac{ \omega}{2} \frac{\dot{\psi}^{2}}{{\psi}^{2}}-3H\omega\frac{\dot{\psi}}{\psi}+ \eta {\phi} ^{2}=0\,, \ \ \ \ \ \ \ \ \ \ \label{eq:EoM-psi}
\end{eqnarray}
where $H=\dot{a}/a$ is the Hubble rate.  Similarly, the variation with respect to the metric gives the following $00$-component
\begin{eqnarray}
{{3}H^{2}\psi}+{{3}H{\dot{\psi}}}-\frac{\omega}{2}\frac{\dot{\psi}^{2}}{\psi}+\eta\phi^{2}\psi-\frac{1}{2}\dot{\phi}^{2}+6\xi H^{2}\phi^{2}+12\xi H \dot{\phi}\phi-{9}\varsigma H^{2}\frac{\dot{\phi}^{2}}{\phi^{2}}=V(\phi)\,, \label{eq:EoM-metric}
\end{eqnarray}
and the variation with respect to $\phi$ renders
\begin{equation}
\ddot{\phi}+3 H \dot{\phi}-12\xi\dot{H}\phi - 24\xi H^{2}\phi+\frac{d V}{d\phi}+6 \varsigma H^{2}\left(\frac{\ddot{\phi}}{\phi^{2}}-\frac{\dot{\phi}^{2}}{\phi^{3}}\right)+18\varsigma {H}^{3}\frac{\dot{\phi}}{\phi^{2}}+
12\varsigma H \dot{H}\frac{\dot{\phi}}{\phi^{2}}-{2}\eta \psi\phi= 0\,.  \label{eq:EoM-phi}
\end{equation}
Let us search for power-law solutions  of the above equations, namely
\begin{equation}
H=\frac{n}{t} \ \ (n>0)\,, \ \ \ \ \ \phi(t)=\phi_{1}\left(\frac{t}{t_{1}}\right)^{\alpha}, \ \ \ \ \ \ \ \ \psi(t)=\psi_{1}\left(\frac{t}{t_{1}}\right)^{\gamma}\,.\label{eq:power law}
\end{equation}
The first expression  is equivalent to $a\propto t^n$.  Here $n,\alpha,\gamma$ are dimensionless parameters and $\phi_{1}$ and $\psi_{1}$ are the  values of the scalar fields at some early time $t=t_1$ in the past when the power-law solutions hold good, e.g. during the EW phase transition.  From its interpretation we must have $\psi_1>0$, and  $|\gamma|\ll1$.  The above scaling solutions are associated to asymptotic states of the different phases of the cosmic evolution,  e.g.  the radiation-dominated ($n=1/2$) and  the matter-dominated ($n=2/3$) epochs,  and inflation ($n\gg 1$)\, \cite{Amendola99,Elahe2017}.


\section{Cosmologically harmless Higgs}

One of the above field equations contains the potential $V$ and the other its derivative $dV/d\phi$. The final expression for  $V$ can be obtained self-consistently by requiring that the field equations determine the same form when we substitute in them the power-law solutions (\ref{eq:power law}). Let us consider arbitrary $n$ values\footnote{For the sake of simplicity, in the original essay I presented the result only for $n=1/2$ (radiation epoch.)}. From Eq.\,(\ref{eq:EoM-psi}), which does not depend on $V$, we find
$\eta \phi_{1}^{2} t^{{2}\alpha}= \left[\omega \gamma \left(\frac{\gamma}{2}+3n-1\right)+{3 n \left(1-{2}n\right)}\right] t_{1}^{{2}\alpha} t^{-2}.$
 It determines the unique value $\alpha=-1$ for the power of $\phi$ in Eq.\,(\ref{eq:power law}).  Using this value,  Eq.\, (\ref{eq:EoM-metric}) leads to
 \small\begin{eqnarray}
V=\frac{\psi_{1}}{ t_{1}^{\gamma}} \left({3n(\gamma+n)}-\frac{\omega}{2} \gamma^{2}+\eta \phi_{1}^{2}t_{1}^{2}\right) t^{\gamma-{2}}+\left(\phi_{1}^{2} t_{1}^{2}\left[6\xi n\left(n-2\right)-\frac{1}{2}\right]-9 n^{2}\varsigma\right) t^{-4}\label{eq:EoM-metric-2}
\end{eqnarray}
and the field equation for $\phi$, Eq.\,(\ref{eq:EoM-phi}),  leads to an alternative expression for $V$:
\begin{eqnarray}
V=-\frac{\psi_{1}}{t_{1}^{\gamma}}\left(\frac{{2}\eta \phi_{1}^{2}t_{1}^{2} }{ \gamma-2}\right)
 t^{\gamma-{2}}-\frac{1}{2}\left[\frac{\phi_{1}^{2}t_{1}^{2}}{2}\left(12\xi n\left(1-2n\right)-3n+2\right)+9 n^2(1-n)  \varsigma \right] t^{-4}\,. \label{eq:EoM-phi-2}
 \end{eqnarray}
In both of them the  two powers  $t^{\gamma-{2}}$ and $t^{-4}$ are consistently present.  When we revert back to $\phi$ through  $t/t_1=\phi_1 /\phi$ (using  $\alpha=-1$ ), these powers become associated to $\phi^{2-\gamma}$ and $\phi^4$, respectively.  Therefore,  it is possible to match the coefficients of each power. We obtain the following unique form for the effective potential (up to an additive constant):
\begin{equation}
V=-\frac{{2}\eta\,\psi_{1}}{\gamma-{2}}\left(\frac{t}{t_1}\right)^{\gamma}\,\phi^{2}+{ \lambda}\phi^{4}\simeq \eta \psi_{1}\, \phi^{{2}}+{\lambda }\phi^{4}\,. \label{eq:potential-phi}
\end{equation}
In  the second equality we used the fact that $|\gamma|\ll1$  and thus the coefficient of $\phi^2$ remains essentially constant. The obtained potential is essentially a Higgs-like potential for $\phi$. The coefficient $\lambda$  of the quartic coupling is given by
\begin{eqnarray}
\lambda=-\frac{3\varsigma n}{\phi_{1}^{4}t_{1}^{4}(12\xi+1)}\,\left[12n\xi \left(n^{2}-n+1\right)+2n-1\right]. \label{eq:Upsilon}
\end{eqnarray}
For instance,  for  $\xi>0$ and  $\varsigma<0$  the necessary condition $\lambda>0$ for vacuum stability is fulfilled both for $n=1/2$ and  $n=2/3$.  The form (\ref{eq:potential-phi}) of the potential tells us that there will be  SSB of the EW symmetry provided $\eta<0$, so this crucial sign is uniquely determined.  Note that  $|\eta|\ll1$ because $\psi_1\sim M_P^2$ and the physical Higgs mass squared is of order  $M_H^2\sim|\eta|\psi_1\sim 10^4$GeV$^4$.  The fact that $|\eta|$ must be a very small parameter has nothing to do with fine-tuning\,\cite{Elahe2017}. As pointed out long ago by Bjorken\,\cite{Bjorken2001a,Bjorken2001b},  it is instead a well-known  feature expected of any theory claiming a possible connection between the gravity scale and the EW scale.

It is also interesting to mention that the consistency conditions leading to the unique expression of the Higgs potential $V$ indicated above  imply a noticeable  relation between the power $\gamma$ that governs the cosmic time evolution of $\psi$ and the BD-parameter $\omega$. Explicit calculation shows that for $\omega\gg n$ it takes the simple form
\begin{equation}\label{eq:gamma2omega}
\gamma^2\omega\simeq{2n}\,.
\end{equation}
It follows that $\gamma$ is naturally small for large $\omega$. A counterpart to this relation for the present universe will be elucidated in the next section, which will provide also a connection between $\omega$ and the (different) power-law evolution of the BD-field in the current epoch.


\section{Late time Universe and effective dynamical dark energy}

As we have seen from the foregoing analysis,  in the generalized BD-gravity context under study the standard solutions $H^2\sim a^{-2/n}$, i.e. $\sim a^{-3}$ and $\sim  a^{-4}$  (for $ n=1/2$ and $2/3$ respectively),  corresponding to matter and radiation,  can perfectly support SSB  with large EW vacuum energy but essentially zero cosmological vacuum energy.   At late time the scalar field $\phi$ becomes suppressed,  $\phi=\phi_1(t_1 /t)\to 0$ (as confirmed from (\ref{eq:power law}) with $\alpha=-1$), and with it $V\to 0$. The field equation for $\phi$,  Eq.\,(\ref{eq:EoM-phi}), completely decouples. The matter component in $S_m$ now becomes relevant and is described as usual by a perfect fluid with proper density  $\rho_m$ and  pressure $p_m$.  Similarly, $\rL$ must also be taken into account at this point.  Around the present  epoch we can just replace the r.h.s. of Eq.\,(\ref{eq:EoM-metric}) by $\rho_m+\rL$ and set $p_m=0$.  Since $\phi\to 0$ quickly with the cosmic evolution, we  recover the standard BD equation  carrying however the remnant  vacuum energy density term, $\rL$, which is necessary  to match the current observation of a CC term:
\begin{eqnarray}\label{eq:FriedmannBD}
3H^{2}+3H\frac{\dot{\psi}}{\psi}-\frac{\omega}{2}\frac{\dot{\psi}^{2}}{\psi^2}=\frac{\rho_m+\rL}{\psi}\,.
\end{eqnarray}
For $\psi=1/(8\pi G)=$const. it reduces to the standard Friedmann's equation of GR.
However,  at this stage of the cosmic evolution we need  also the $ii$-component of the BD field equations emerging from variating the action with respect to the metric, as this equation is linked to the pressure components. While $p_m$ can be neglected,  $p_{\Lambda}=-\rL$ cannot be neglected anymore.  The relevant pressure equation reads
\begin{equation}
2\dot{H} + 3H^2 + \frac{\ddot{\psi}}{\psi} + 2H\frac{\dot{\psi}}{\psi} + \frac{\omega}{2}\frac{\dot{\psi}^2}{\psi^2} = -\frac{1}{\psi}p_{\Lambda}\,.
\label{pressureBD}
\end{equation}
Correspondingly, for $\psi=1/(8\pi G)=$const. it boils down to the standard pressure equation in GR. However, $\psi$  is not exactly constant in our  time. It still evolves mildly with the expansion, albeit at a slightly different rate that we can estimate.  Once more we seek power-law solutions. This time we use the form
\begin{equation}\label{eq:powernu}
\psi=\psi_{0}\, a^{-\epsilon}=M_P^2\, a^{-\epsilon}\,,
\end{equation}
since the scale factor can easily be related to redshift ($z=a^{-1}-1$) and hence to observations. Here we have normalized the scale factor at present to $a_0=1$.  The power $\epsilon$  is of course small in absolute value and is the analogue in the modern epoch of the power $\gamma$ used earlier.  The relevant BD field equations for the current epoch are the following: Eq.\,(\ref{eq:EoM-psi}) -- dropping $\phi$ in it --  (\ref{eq:FriedmannBD}) and (\ref{pressureBD}).   Substituting the ansatz (\ref{eq:powernu}) in (\ref{eq:FriedmannBD})
and using the fact that  $|\epsilon|\ll 1$ we can  put the result
in an approximate  $\CC$CDM-like form,  namely
\begin{equation}\label{eq:FriedmannDDE}
H^2=\frac{8\pi G}{3}\left(\rho_{m 0} a^{-3+\epsilon}+\rD(H)\right)\,,
\end{equation}
with
\begin{equation}\label{eq:rLeff}
  \rD(H)=\rL+\frac{3\nu}{8\pi G} H^2\,.
\end{equation}
Here we have defined  $\nu\equiv \epsilon(1+\omega\epsilon/6)$. Recall that $1/G\equiv 1/G(t_0)=8\pi\psi(t_0)$.
Remarkably, the expression  (\ref{eq:rLeff})  emulates an effective dynamical dark energy  (DDE), which  is  induced by the slow dynamics of the   BD field.  The  present physical value is $\rLo\equiv\rD(H_0)=\rL+{3\nu}/{(8\pi G}) H_0^2$, very close to the $\rL$-term in the original action (since $|\epsilon|\ll 1$ and hence $|\nu|\ll 1$ ).  The anomalous matter conservation law  $\rho_m=\rho_{m 0} a^{-3+\epsilon}$ appearing in (\ref{eq:FriedmannDDE})  mimics  an interaction of matter with the DDE. For $\epsilon=0$ (hence $\nu=0$) we recover the exact situation of the $\CC$CDM. Worth noticing, for $\epsilon\neq 0$  the expression (\ref{eq:rLeff}) adopts  the form of the running vacuum model (RVM), see \cite{JSPReview2013,Fossil2008,BarLambdaCDM2015} and references therein.  In Ref.\,\cite{ApJpapers,EPL2017,EPL2018,MNRAS2018a,MNRAS2018b} it was shown that the RVM fits the overall observations better  than the $\CC$CDM.

The product $\omega\epsilon$ becomes determined by the remaining two BD-equations (\ref{eq:EoM-psi}) and (\ref{pressureBD}) upon using the ansatz (\ref {eq:powernu}).  After a detailed calculation and  evaluation at $a=1$  (redshift $z=0$) one finds
\begin{equation}\label{eq:omegatimesepsilon}
\omega\epsilon=-\frac{4 - 3\Omega_m }{2-\Omega_m }+{\cal O}(\epsilon)\simeq -1.8\,\ \ \ \ \ \ (\textrm{for}\ \Omega_m\simeq 0.3)\,.
\end{equation}
Notice that $|\omega\epsilon|\gg\epsilon$.
Moreover, $\omega\epsilon^2\sim\epsilon$ and the two terms defining $\nu=\epsilon+\omega\epsilon^2/6$ are indeed of the same order.  The $\sim \omega\epsilon^2$ part of (\ref{eq:rLeff})  is easily seen to originate from the non-canonical kinetic term of the BD-field, cf. Eq.\,(\ref{eq:BDaction}), interpreted roughly in the manner of an effective potential contribution to the vacuum dynamics:
\begin{equation}\label{eq:noncanonicalKineticE}
 -\frac{\omega}{\psi}g^{\mu\nu}\partial_{\nu}\psi\partial_{\mu}\psi\ \longrightarrow\  \frac{\omega}{\psi}\dot\psi^2\sim \omega \epsilon^2 \psi H^2\sim \omega\epsilon^2 M_P^2 H^2\,,
\end{equation}
where we have used (\ref{eq:powernu}).  The precise and rigorous contribution, however, appears at the level of the field equations and stems from the third term on the \textit{l.h.s} of Eq.\,(\ref{eq:FriedmannBD}). Together with the second term in that equation (which is linear in $\epsilon$ within the context of the power-law solution we are considering) it leads to the running vacuum form (\ref{eq:FriedmannBD}) \cite{JSPReview2013,Fossil2008,BarLambdaCDM2015}.  On comparing (\ref{eq:gamma2omega}) with (\ref{eq:omegatimesepsilon}) we can see that in both cases the power-law evolution of the BD-field becomes progressively  smaller (hence closer to a constant) for larger and larger values of  $\omega$, but at different rates:  for the current epoch  $\epsilon\sim 1/\omega$, whereas in the early universe $\gamma\sim 1/\sqrt{\omega}$.

The  above cosmological solution is obviously different from GR and has distinct properties.  The pressure equation (\ref{pressureBD}) can also be put in approximate $\CC$CDM fashion using (\ref{eq:powernu}).  Neglecting the small, ${\cal O}(\epsilon)$, additive corrections to the parameters except for those in the dynamical terms, such as $\sim  a^{-3+\epsilon}$ and $\sim \nu M_P^2H^2$,  we find the leading expression for the effective acceleration equation:
\begin{equation}\label{eq:currentacceleration}
\frac{\ddot{a}}{a}=-\frac{4\pi G}{3}\,\left(\rho_m^0 a^{-3+\epsilon}+\rD(H)+3p_{\Lambda}\right)\,.
\end{equation}
It follows that the equation of state (EoS) for the effective DDE reads
\begin{equation}\label{eq:EffEoS}
  w=\frac{p_{\Lambda}}{\rD(H)}\simeq -1+\frac{3\nu}{8\pi G \rL}\,H^2=-1+\frac{\nu}{\Omega_{\Lambda}}\,\frac{H^2}{H_0^2}\,.
\end{equation}
Thus, for $\epsilon>0\ (\epsilon<0) $  we have $\nu>0\  (\nu<0)$ and the effective DDE behaves quintessence (phantom)-like.  For $\epsilon=0$ (i.e. $\nu=0$) we have  $w=-1$ ($\CC$CDM) and only then the BD-parameter is forced to $|\omega|\to\infty$ from (\ref{eq:omegatimesepsilon}). For a phenomenological confrontation of this class of BD solutions with the current observations, see \cite{JavierJoan2018}.

Needless to say, matter is locally and covariantly conserved in BD-theory, as there is actually no interaction of matter with the BD-field $\psi$. However, when we try to encapsulate the observational behavior of BD-gravity  in a strict GR-form, we find that the former may effectively appear as a model of interactive quintessence or phantom DE.  The eventual detection of this kind of ``anomalies'' in the behavior of the $\CC$CDM  could be signaling the possible presence of BD-gravity at a more fundamental level.  This does not preclude, of course, that other sources of fundamental DE dynamics associated to new fields might also be concomitant with the same physical phenomenon, see e.g. \cite{BanerjeePavon01}. However, BD-gravity alone does indeed have the capacity to mimic dynamical DE.

The following remark is in order. Note that if the tiny time dependence of the coefficient of $\phi^2$ in the first expression of the potential in Eq.\,(\ref{eq:potential-phi}) had not been neglected, the expectation value of the  Higgs field derived from BD-gravity  would carry a mild time-evolution, $\langle\phi\rangle \sim t^{\gamma/2}$\, ($|\gamma|\ll1$), which would be transferred to all of the particle masses that are generated through SSB, a conclusion that can also  be reached  from alternative considerations\,\cite{FritzschSola,Terazawa}. Thus the BD-framework under study naturally  leads to a time variation of the fundamental constants such as particle masses and couplings (e.g. Fermi's effective coupling $G_F$, which goes like  the inverse of the  $W$ mass squared).

\section{Discussion}

Let us close with a summary and a few additional considerations.  In this paper, we have shown that the BD-gravity framework can be appropriately extended by introducing  a new scalar field $\phi$ (different from the BD-field $\psi$)  which is also nonminimally  coupled to gravity. The effective potential for $\phi$, $V(\phi)$, can be generated self-consistently by searching  for cosmologically viable solutions of the coupled  field equations for the two scalar fields and the gravitational field.  The structure for $V(\phi)$  adopts the general form of the Higgs potential, and for this reason it  is tempting to identify $\phi$ with the Higgs field. If that construction would be fully consistent (what is currently beyond the scope of this essay)  it could be a way to motivate the Higgs potential from a gravity framework,  specifically from scalar-tensor theories, a property which for the time being does not seem to be at reach from pure GR.  Additional studies should address the problem of stability of the power-law  solutions that have been found and consider them in a thermodynamical context, more appropriate for the early universe. The decay evolution of the scalar field $\phi$ and of the corresponding potential should lead to the extinction of the huge vacuum energy density, in a manner similar to the one that has been described here.  The scalar-tensor framework considered in this work permits the separation of the cosmological vacuum energy density from  the QFT vacuum energy density, and in fact the latter can be huge in the early universe while still coexisting with a vanishingly small value of the former, as shown by the fact that the Hubble function involved in the obtained solution has no trace of vacuum  energy from $V(\phi)$.  Finally, let us note that the CC problem is really a multifarious problem, and it cannot be narrowed down to just the  SSB  contribution associated to the Higgs mechanism.  In particular,  the Higgs sector of the standard model of particle physics suffers from the problem of naturalness since it is unprotected from the  quadratic divergences induced on the value of the  Higss mass when embedding the EW Theory into a Grand Unified Theory.  Therefore, even if dodging around  the huge contribution from the Higgs potential to the CC, the Higgs sector itself needs a deeper substantiation within pure particle physics. And if that is not enough, the scope of the CC problem really goes beyond the Higgs and the entire EW sector, as many other contributions are possible that can further aggravate the problem. Definitely,  more study is required to solve these fundamental conundrums. Here, we have hinted at the existence of a new road towards alleviating some of them in a context beyond GR, but not too far from it.

\section{Conclusions}

Generalized BD-gravity theories admit cosmologically viable solutions which lead to the precise structure of the Higgs potential.
The  inherent QFT vacuum energy of the potential  is as large as usual but the cosmological vacuum energy density remains very small. The latter emerges only at late times through the usual constant term $\rL$, but accompanied with an additional (albeit dynamical) contribution induced by the slow time evolution of the BD-field $\psi$.  Effectively it  leads to  $\CC(t)$CDM, with a moderate time-evolving $\CC(t)=\CC(H(t))$ and EoS very close to $-1$.  It mimics the running vacuum model, which has recently been shown to provide a  fit to the overall cosmological data better than the $\CC$CDM. Finding mild vacuum dynamics, accompanied with unsuspected time-evolving particle masses,  could be the ``smoking gun'' signaling that the underlying gravity theory is not GR but BD. It could provide the missing link between gravity and particle physics.




\vspace{0.5cm}


{\bf Acknowledgments}

\noindent I am supported by  MINECO FPA2016-76005-C2-1-P,  2017-SGR-929 (Generalitat
de Catalunya) and MDM-2014-0369 (ICCUB).  I thank the (anonymous) referee  for  helpful and constructive comments, which have helped me to improve the presentation of the main ideas involved in this work.


\end{document}